# A unified multirate lattice Boltzmann framework for thermosolutal dendritic solidification

Yang Liu[a], Xiaomin Wu[a,*] and Chengjie Zhan[b,*]

[a] Key Laboratory for Thermal Science and Power Engineering of Ministry of Education, Department of Energy and Power Engineering, Tsinghua University, Beijing 100084, China
[b] Department of Engineering Mechanics, Tsinghua University, Beijing 100084, China
Corresponding authors: wuxiaomin@mail.tsinghua.edu.cn; zhancj@mail.tsinghua.edu.cn

ABSTRACT

Thermosolutal dendritic solidification involves interface evolution, solute diffusion, heat transfer, and melt flow over markedly different time scales. In lattice Boltzmann simulations, a single numerical time interval may place different transport processes in unfavorable relaxation ranges, while asynchronous updates require consistent transfer of phase-change contributions. To address these issues, a unified multirate multiple-relaxation-time lattice Boltzmann method is developed for thermal, solutal, and thermosolutal dendritic solidification. The coupled fields share a common moment-space framework but evolve at different update rates. The concentration and temperature source terms are separated into transport-related and phase-change contributions, and each resolved phase increment is used to immediately transfer the corresponding solutal and latent-heat contributions. The method reproduces characteristic dendritic morphologies and tip-velocity trends under pure diffusion and forced convection, with weak sensitivity to the tested update factors. Directional solidification over Lewis numbers $Le = 1$–$1000$ captures the transition from nearly planar to cellular and strongly branched growth, including vertically aligned dendrites and solute-rich interdendritic channels for saline water, in qualitative agreement with experiments. Source-coupling ablation shows that delayed coarse-step transfer produces increasingly strong local source pulses and eventual loss of numerical stability as time-scale separation increases, whereas phase-step instantaneous transfer remains stable over the tested conditions. These results demonstrate the applicability of the proposed framework to dendritic solidification with strongly separated transport time scales. The source code is publicly available in the *DendriteLBM* repository.

## 1. Introduction

Dendritic solidification is a canonical interfacial pattern-forming process and an important source of microstructural heterogeneity in solidifying materials [1–3]. In metallic alloys, dendrite orientation, characteristic spacing, and solute redistribution strongly influence grain competition and microsegregation [4]. Similar interfacial instabilities arise during directional freezing of aqueous solutions, where solute rejection promotes the formation of cellular or dendritic ice structures together with brine pockets and solute-rich channels [5–7]. Although these systems differ substantially in material properties and characteristic scales, their mesoscale evolution involves coupled anisotropic interface motion, heat transfer, solute redistribution, and melt flow. Quantitative simulation therefore requires not only an accurate representation of the evolving solid-liquid interface, but also an appropriate treatment of transport processes with strongly separated time scales.

The phase-field method (PFM) provides a natural framework for such problems by representing the moving solid-liquid interface with a continuous order parameter, thereby avoiding explicit interface tracking [8]. Based on thin-interface analysis, Karma showed that the finite numerical interface width in quantitative alloy phase-field models can introduce nonphysical effects such as artificial solute trapping, and proposed the anti-trapping current to compensate for these effects [9, 10]. For nonisothermal binary solidification, coupled heat and solute transport must additionally be considered to recover the intended sharp-interface behavior [11].

Melt convection further increases the coupling complexity between interface evolution and transport in the surrounding liquid [13]. The lattice Boltzmann method (LBM) has been widely used for fluid-flow simulations because of its local collision-streaming procedure and straightforward parallel implementation [14]. Its kinetic formulation can also be extended to convection-diffusion and other scalar transport equations, making it well suited for multiphysics problems in which momentum and scalar fields are solved within a common numerical framework. Multiple-relaxation-time (MRT) collision operators further allow different kinetic modes to be adjusted independently, thereby improving numerical stability [15]. On this basis, PFM-LBM formulations were introduced for dendritic-growth simulations with melt convection [16]; subsequent developments of anisotropic LBM formulations enabled the phase-field equation itself to be solved within the LBM framework [17, 18]. Related studies then progressed toward unified treatment of multiple coupled fields, with the phase, concentration, temperature, and flow fields incorporated within a common MRT-LBM framework [19–22]. More recently, these approaches have been extended to more complex solidification processes involving dendrite motion, sedimentation, and bubble coupling [23–25], as well as to freezing desalination and sea-ice crystallization [26–28]. Combined with GPU parallelization and adaptive-mesh techniques, PFM-LBM has also shown good potential for large-scale dendritic simulations [29, 30]. Overall, these studies demonstrate the flexibility of LBM for multiphysics dendritic solidification and provide a methodological basis for unified modeling of coupled thermal, solutal, and hydrodynamic processes.

Despite these developments, temporal discretization across coupled fields remains an important numerical issue. In LBM, the relaxation parameters depend on both the physical transport coefficients and the numerical time intervals. In thermosolutal solidification, the thermal diffusivity can exceed the solutal diffusivity by several orders of magnitude, leading to strongly separated transport time scales, particularly at large Lewis numbers $Le = \alpha_L/D_L$ [12]. If interfacial dynamics, solutal diffusion, thermal diffusion, and hydrodynamic transport are all advanced with the same time step, an unfavorable compromise may arise among numerical stability, temporal resolution, and computational cost. To address such disparate time scales, Dorari et al. [31] proposed a multiple-grid-time-step LBM that allows different transport processes to evolve on independent numerical scales. Li [32] further developed a multiple-time-scaling strategy for temporal decoupling of the convection-diffusion equation (CDE). Zhang et al. [33] employed a hierarchical PFM-LBM algorithm to handle disparate physical time scales in large-scale thermosolutal simulations with $Le$ approaching $10^4$. Wang et al. [21] further assigned different time steps to the phase, concentration, temperature, and flow distributions and reported only weak sensitivity of dendrite-tip kinetics to the selected scaling factors. These studies demonstrate that multirate evolution can alleviate the numerical constraints associated with strongly separated transport time scales.

Assigning independent update intervals to different fields, however, does not by itself determine how coupling quantities should be transferred between asynchronously advanced fields. More generally, multirate multiphysics simulations require temporally consistent transfer of coupling quantities between fields advanced on different update schedules [34, 35]. In dendritic solidification, latent-heat release and solute redistribution, including the anti-trapping correction, are directly tied to the actual phase evolution [10]. When the temperature or concentration field is updated less frequently than the phase field, the temporal discretization and transfer of these phase-change contributions become particularly important. A multirate coupling strategy is therefore needed to transfer the generated thermal and solutal contributions consistently with the resolved interface evolution.

In this work, a unified multirate MRT-LBM is developed for dendritic solidification with coupled thermal, solutal, and hydrodynamic effects. The coupled fields share a common moment-space framework while evolving at different update rates, allowing the numerical time scales of individual transport processes to be adjusted independently. To maintain consistent coupling under asynchronous evolution, a phase-increment-based source treatment is introduced. At each phase-field update, the resolved phase increment is used to construct and immediately transfer the corresponding latent-heat and solutal contributions, including the anti-trapping contribution, while the scalar transport operators retain their own update schedules. The proposed framework is validated against thermal, solutal, and thermosolutal dendritic benchmarks, and is further examined through directional saline-water solidification and controlled Lewis-number variations over $Le = 1$–$1000$. A source-coupling ablation is additionally performed to assess the effect of delayed coarse-step accumulation of phase-change contributions on numerical stability.

## 2. Macroscopic governing equations

### 2.1. Phase field

To model dendritic growth, the evolution of the solid-liquid interface is described by the phase-field variable $\phi$. The value $\phi = 1$ denotes the solid phase, $\phi = -1$ denotes the liquid phase, and $-1 < \phi < 1$ represents the diffuse solid-liquid interface. The local liquid fraction is therefore given by $f_L = (1 - \phi)/2$. Following the quantitative thermosolutal phase-field formulations for dendritic solidification [11, 20, 21], the phase-field equation is written as

$$a_s^2(\mathbf{n})\tau_0 \frac{\partial \phi}{\partial t} = W_0^2 \nabla \cdot \left[a_s^2(\mathbf{n}) \nabla \phi\right] + W_0^2 \nabla \cdot \mathbf{N} + Q_\phi, \tag{1}$$

where $a_s(\mathbf{n})$ is the anisotropic interface function, $\tau_0$ is the phase-field time scale, $W_0$ is the characteristic interface width, $\mathbf{N}$ is the anisotropic correction vector, and $Q_\phi$ is the phase-transition source term.

To describe the preferred growth directions of dendrites in the present two-dimensional simulations, the fourfold anisotropy function is adopted as [17, 18]

$$a_s(\mathbf{n}) = 1 - 3\epsilon_s + 4\epsilon_s \left(n_x^4 + n_y^4\right), \tag{2}$$

where $\epsilon_s$ is the anisotropic strength. The interfacial unit normal vector $\mathbf{n} = (n_x, n_y)^\top$ is calculated from the phase-field gradient as $\mathbf{n} = -\nabla\phi/|\nabla\phi|$. The correction vector $\mathbf{N}$ accounts for anisotropic surface energy and is defined as

$$\mathbf{N} = |\nabla\phi|^2 a_s(\mathbf{n}) \left[\frac{\partial a_s(\mathbf{n})}{\partial(\partial_x\phi)}, \frac{\partial a_s(\mathbf{n})}{\partial(\partial_y\phi)}\right]^\top. \tag{3}$$

The source term $Q_\phi$ is associated with the double-well potential and the local thermosolutal driving force for dendritic growth, which is expressed as

$$Q_\phi = \phi(1-\phi^2) - \lambda(M_c U + \theta)(1-\phi^2)^2, \tag{4}$$

where $\lambda$ is the coupling coefficient. Following the thermosolutal phase-field formulation [11, 21], $U$, $\theta$, and $M_c$ denote the dimensionless concentration (supersaturation), temperature (undercooling), and solutal coupling parameter, respectively, and are defined as

$$\begin{aligned} U &= \frac{\dfrac{2C/C_\infty}{(1+k_c)-(1-k_c)\phi} - 1}{1-k_c}, \\ \theta &= \frac{T - T_m - m_\mathrm{L} C_\infty}{L_h/C_{p,\mathrm{L}}}, \\ M_c &= -\frac{m_\mathrm{L}(1-k_c)C_\infty}{L_h/C_{p,\mathrm{L}}}. \end{aligned} \tag{5}$$

where $C$ and $T$ denote the concentration and temperature fields, respectively; $C_\infty$ is the far-field concentration, i.e., the initial concentration of the solution; $k_c = C_\mathrm{S}/C_\mathrm{L}$ is the partition coefficient, i.e., the ratio of equilibrium solute concentrations at the solid-liquid interface; $m_\mathrm{L}$ is the slope of the liquidus line in the phase diagram, and $T_m$, $L_h$, and $C_{p,\mathrm{L}}$ are the melting temperature, latent heat, and specific heat capacity, respectively.

## 2.2. Concentration field

The solutal transport is described using the quantitative phase-field formulation with anti-trapping correction [9–11, 21]. The dimensionless concentration $U$ satisfies

$$\frac{\partial U}{\partial t} + \mathbf{u}\cdot\nabla U = \nabla\cdot(D_\mathrm{eff}\nabla U) + Q_U, \tag{6}$$

where $\mathbf{u} = (u_x, u_y)^\top$ is the melt velocity obtained from the flow field, and $D_\mathrm{eff}$ is the phase-dependent effective solutal diffusivity, explicitly written as

$$D_\mathrm{eff} = \frac{(1+\phi)D_\mathrm{S} + (1-\phi)D_\mathrm{L}}{(1+k_c)-(1-k_c)\phi}, \tag{7}$$

where $D_\mathrm{L}$ and $D_\mathrm{S}$ are the solutal diffusivities in the liquid and solid phases, respectively.

The source term $Q_U$ can be written as

$$Q_U = -\frac{(1-k_c)D_\mathrm{eff}}{(1+k_c)-(1-k_c)\phi}\nabla U\cdot\nabla\phi + \frac{\left[1+(1-k_c)U\right]\partial_t\phi - 2\nabla\cdot\mathbf{j}_\mathrm{at}}{(1+k_c)-(1-k_c)\phi}, \tag{8}$$

where the first term represents the correction associated with the phase-dependent diffusion coefficient, while the second term accounts for solute redistribution induced by interface motion. The anti-trapping current $\mathbf{j}_\mathrm{at}$ introduced in quantitative alloy phase-field models [9, 10] is written as

$$\mathbf{j}_\mathrm{at} = \frac{W_0\partial_t\phi}{2\sqrt{2}}\left[1+(1-k_c)U\right]\mathbf{n}. \tag{9}$$

### 2.3. Temperature field

Heat transfer is described by a CDE for the temperature field $T$ with a source term $Q_T$. Unlike the constant-property thermal formulations commonly adopted in previous models [20–22], the present formulation accounts for solid-liquid differences in thermal conductivity and volumetric heat capacity. The governing equation is [36, 37]

$$\frac{\partial T}{\partial t} + \mathbf{u} \cdot \nabla T = \nabla \cdot (\alpha_{\mathrm{eff}} \nabla T) + Q_T, \tag{10}$$

where $\alpha_{\mathrm{eff}} = \kappa_{\mathrm{eff}}/(\rho C_p)_{\mathrm{eff}}$ is the phase-dependent effective thermal diffusivity. The effective thermal conductivity $\kappa_{\mathrm{eff}}$ and volumetric heat capacity $(\rho C_p)_{\mathrm{eff}}$ are interpolated by the phase-field variable $\phi$ as

$$\begin{aligned} \kappa_{\mathrm{eff}} &= \frac{1+\phi}{2}\kappa_{\mathrm{S}} + \frac{1-\phi}{2}\kappa_{\mathrm{L}}, \\ (\rho C_p)_{\mathrm{eff}} &= \frac{1+\phi}{2}\rho_{\mathrm{S}} C_{p,\mathrm{S}} + \frac{1-\phi}{2}\rho_{\mathrm{L}} C_{p,\mathrm{L}}, \end{aligned} \tag{11}$$

where the upright subscripts “S” and “L” denote the solid- and liquid-phase properties, respectively.

The source term $Q_T$ consists of two parts, written as

$$Q_T = \frac{\rho_{\mathrm{S}} C_{p,\mathrm{S}} - \rho_{\mathrm{L}} C_{p,\mathrm{L}}}{2(\rho C_p)_{\mathrm{eff}}} \alpha_{\mathrm{eff}} \nabla T \cdot \nabla \phi + \frac{\rho_{\mathrm{L}} L_h \partial_t \phi}{2(\rho C_p)_{\mathrm{eff}}}, \tag{12}$$

where the first term compensates for the spatial variation of the effective volumetric heat capacity, so that the thermal diffusion operator recovers $\nabla \cdot (\kappa_{\mathrm{eff}} \nabla T)/(\rho C_p)_{\mathrm{eff}}$ [36–38]; the second term represents latent-heat release associated with phase change.

### 2.4. Flow field

The melt flow is assumed to be incompressible and Newtonian and is governed by the Navier-Stokes equations, written as

$$\begin{aligned} &\nabla \cdot \mathbf{u} = 0, \\ &\frac{\partial \mathbf{u}}{\partial t} + \mathbf{u} \cdot \nabla \mathbf{u} = -\frac{\nabla p}{\rho_0} + \nu \nabla^2 \mathbf{u} + \mathbf{F}_b, \end{aligned} \tag{13}$$

where $\rho_0$ is the reference density, taken as the initial liquid density $\rho_{\mathrm{L}}$; $\nu$ is the kinematic viscosity of the liquid phase, and $\mathbf{F}_b$ is the body force term.

## 3. Unified MRT-LBM and multirate strategy

### 3.1. D2Q9-based MRT collision operator

A common MRT-LBM framework based on the two-dimensional nine-velocity (D2Q9) lattice is employed for the four coupled fields. The scalar CDEs are discretized mainly following the MRT-LBM formulation for CDE of Zhang et al. [39], while the common moment-space construction follows Chai and Shi [40]. Within this framework, the phase field [$\phi$, Eq. (1)], concentration field [$U$, Eq. (6)], temperature field [$T$, Eq. (10)], and flow field [$F$, Eq. (13)] share the same moment-space collision structure. The generic collision step is written as

$$f_i^{\chi,\dagger}(\mathbf{x},t) = f_i^{\chi}(\mathbf{x},t) - \mathbf{M}^{-1}\mathbf{S}^{\chi}\left(\mathbf{m}^{\chi} - \mathbf{m}^{\chi,\mathrm{eq}}\right) + \mathbf{M}^{-1}\left(\mathbf{I} - \frac{\mathbf{S}^{\chi}}{2}\right)\mathbf{G}_F + \delta t_{\chi} \mathbf{\Phi}_{\chi}, \tag{14}$$

where $\chi \in \{\phi, U, T, F\}$ denotes a generic field; $f_i^{\chi}$ is the distribution function in the $i$-th discrete direction at position $\mathbf{x}$ and time $t$, and $f_i^{\chi,\dagger}$ denotes its post-collision value. The corresponding moment vectors are defined as $\mathbf{m}^{\chi} = \mathbf{M}\mathbf{f}^{\chi}$, and $\mathbf{m}^{\chi,\mathrm{eq}}$ indicates the equilibrium moment. $\mathbf{M}$ is the transformation matrix and $\mathbf{S}^{\chi}$ is the field-specific diagonal relaxation matrix. The forcing vector $\mathbf{G}_F$ is included only for the flow field, whereas for the three scalar fields, the $i$-th component of the discrete source vector is given by $\Phi_{\chi,i} = \omega_i Q_{\chi}$. The scalar source treatment follows the MRT-LBM formulation for CDE [39], while the flow forcing term is incorporated using Guo’s scheme [41].

The classical D2Q9 lattice and moment basis [15, 42] are used for all four fields. The discrete velocity set $\mathbf{e}_i$ is defined as

$$\mathbf{e}_i = \begin{cases} (0,0), & i = 0, \\ (1,0),(0,1),(-1,0),(0,-1), & i = 1,2,3,4, \\ (1,1),(-1,1),(-1,-1),(1,-1), & i = 5,6,7,8. \end{cases} \tag{15}$$

The D2Q9 lattice weights are $\omega_0 = 4/9$, $\omega_{1-4} = 1/9$, and $\omega_{5-8} = 1/36$, and the transformation matrix $\mathbf{M}$ is

$$\mathbf{M} = \begin{bmatrix} 1 & 1 & 1 & 1 & 1 & 1 & 1 & 1 & 1 \\ -4 & -1 & -1 & -1 & -1 & 2 & 2 & 2 & 2 \\ 4 & -2 & -2 & -2 & -2 & 1 & 1 & 1 & 1 \\ 0 & 1 & 0 & -1 & 0 & 1 & -1 & -1 & 1 \\ 0 & -2 & 0 & 2 & 0 & 1 & -1 & -1 & 1 \\ 0 & 0 & 1 & 0 & -1 & 1 & 1 & -1 & -1 \\ 0 & 0 & -2 & 0 & 2 & 1 & 1 & -1 & -1 \\ 0 & 1 & -1 & 1 & -1 & 0 & 0 & 0 & 0 \\ 0 & 0 & 0 & 0 & 0 & 1 & -1 & 1 & -1 \end{bmatrix}. \tag{16}$$

For convenience, the lattice spacing is fixed at $\delta x = 1$, whereas the lattice time interval $\delta t_\chi$ is allowed to differ among the four fields under the multirate strategy. The lattice speed and sound speed associated with field $\chi$ are

$$c_\chi = \frac{\delta x}{\delta t_\chi}, \qquad c_{\mathrm{s},\chi}^2 = \frac{c_\chi^2}{3}. \tag{17}$$

Accordingly, the discrete lattice velocities are $\mathbf{c}_i^\chi = c_\chi \mathbf{e}_i$. The field-specific time interval also enters the relation between the relaxation rates in $\mathbf{S}^\chi = \mathrm{diag}(s_0^\chi, s_1^\chi, \dots, s_8^\chi)$ and the corresponding macroscopic transport coefficients, as detailed in Section 3.2.

## 3.2. Moment-space evolution details for the four fields

Based on the common MRT operator introduced in Section 3.1, four sets of distribution functions are employed for the phase field ($f_i^\phi$), concentration field ($f_i^U$), temperature field ($f_i^T$), and flow field ($f_i^F$). The scalar-field discretization follows the MRT-LBM formulations [39, 40], while field-specific numerical time intervals are introduced within the multirate treatment [32]. Although all four fields share the same D2Q9 lattice and moment transformation matrix, their equilibrium moments, relaxation matrices, source or forcing terms, and streaming procedures are specified according to the corresponding macroscopic equations.

### *3.2.1. Scalar-field evolution*

The three scalar fields, namely the phase-field variable $\phi$, the dimensionless concentration $U$, and the temperature $T$, are recovered from their respective distribution functions as

$$\phi = \sum_i f_i^\phi, \qquad U = \sum_i f_i^U, \qquad T = \sum_i f_i^T. \tag{18}$$

For the concentration and temperature fields, the equilibrium distributions follow the standard linear convection-diffusion form, i.e.,

$$f_i^{U,\mathrm{eq}} = \omega_i U \left(1 + \frac{3\mathbf{e}_i \cdot \mathbf{u}}{c_U}\right), \qquad f_i^{T,\mathrm{eq}} = \omega_i T \left(1 + \frac{3\mathbf{e}_i \cdot \mathbf{u}}{c_T}\right), \tag{19}$$

where $\mathbf{u}$ is the melt velocity. The corresponding equilibrium moments are

$$\begin{aligned} \mathbf{m}^{U,\mathrm{eq}} &= \left(U, -2U, U, \frac{u_x U}{c_U}, -\frac{u_x U}{c_U}, \frac{u_y U}{c_U}, -\frac{u_y U}{c_U}, 0, 0\right)^\mathsf{T}, \\ \mathbf{m}^{T,\mathrm{eq}} &= \left(T, -2T, T, \frac{u_x T}{c_T}, -\frac{u_x T}{c_T}, \frac{u_y T}{c_T}, -\frac{u_y T}{c_T}, 0, 0\right)^\mathsf{T}. \end{aligned} \tag{20}$$

For the phase field, the same moment-space structure is combined with the anisotropic LBM treatment developed for dendritic PFM [17, 18, 21]. The equilibrium distribution function is constructed as

$$f_i^{\phi,\mathrm{eq}} = \omega_i \left(\phi + \frac{3\mathbf{e}_i \cdot \mathbf{v}_\mathrm{n}}{c_\phi}\right), \tag{21}$$

where $\mathbf{v}_\mathrm{n}$ is an auxiliary anisotropic advection vector associated with the correction term $\nabla \cdot \mathbf{N}$ in Eq. (1), defined as $\mathbf{v}_\mathrm{n} = -W_0^2\mathbf{N}/\tau_0$. Explicitly, the equilibrium moment vector is

$$\mathbf{m}^{\phi,\mathrm{eq}} = \left(\phi, -2\phi, \phi, \frac{v_{\mathrm{n},x}}{c_\phi}, -\frac{v_{\mathrm{n},x}}{c_\phi}, \frac{v_{\mathrm{n},y}}{c_\phi}, -\frac{v_{\mathrm{n},y}}{c_\phi}, 0, 0\right)^\mathsf{T}. \tag{22}$$

The three scalar fields employ the same diagonal relaxation structure, i.e.,

$$\mathbf{S}^\chi = \mathrm{diag}\left(1, 1, 1, s_3^\chi, 1, s_5^\chi, 1, 1, 1\right), \qquad \chi \in \{\phi, U, T\}, \tag{23}$$

where $s_3^\chi = s_5^\chi = 1/\tau_\chi$ control the diffusive flux moments. Consistent with the transport-coefficient relation recovered by Zhang et al. [39], and allowing field-specific numerical time intervals within the multirate strategy [32], the corresponding relaxation times are

$$\tau_\phi = 3\delta t_\phi \frac{a_\mathrm{s}^2(\mathbf{n})W_0^2}{\tau_0} + \frac{1}{2}, \qquad \tau_U = 3\delta t_U D_\mathrm{eff} + \frac{1}{2}, \qquad \tau_T = 3\delta t_T \alpha_\mathrm{eff} + \frac{1}{2}. \tag{24}$$

The concentration and temperature distributions follow the standard streaming step as

$$\begin{aligned} f_i^U(\mathbf{x} + \mathbf{e}_i\delta x, t + \delta t_U) &= f_i^{U,\dagger}(\mathbf{x}, t), \\ f_i^T(\mathbf{x} + \mathbf{e}_i\delta x, t + \delta t_T) &= f_i^{T,\dagger}(\mathbf{x}, t). \end{aligned} \tag{25}$$

Particularly for the phase field, an anisotropic streaming treatment is additionally applied to recover the anisotropic phase-field equation [17, 18, 21], i.e.,

$$a_\mathrm{s}^2(\mathbf{n}) f_i^\phi(\mathbf{x} + \mathbf{e}_i\delta x, t + \delta t_\phi) = f_i^{\phi,\dagger}(\mathbf{x}, t) - \left[1 - a_\mathrm{s}^2(\mathbf{n})\right] f_i^\phi(\mathbf{x} + \mathbf{e}_i\delta x, t). \tag{26}$$

#### *3.2.2. Flow-field evolution*

The melt flow is solved using the standard D2Q9 MRT-LBM formulation [15, 40]. The fluid density and velocity are recovered as

$$\rho = \sum_i f_i^F, \qquad \rho\mathbf{u} = \sum_i f_i^F c_F \mathbf{e}_i + \frac{\delta t_F}{2}\mathbf{F}_b. \tag{27}$$

Under the weakly compressible approximation, density variations remain small and $\rho$ is treated as approximately constant. The equilibrium distribution function of the flow field is given by

$$f_i^{F,\mathrm{eq}} = \omega_i \rho \left[1 + \frac{3\mathbf{e}_i \cdot \mathbf{u}}{c_F} + \frac{9(\mathbf{e}_i \cdot \mathbf{u})^2}{2c_F^2} - \frac{3\mathbf{u}^2}{2c_F^2}\right]. \tag{28}$$

The corresponding equilibrium moment vector is

$$\mathbf{m}^{F,\mathrm{eq}} = \rho\left(1, -2 + \frac{3\mathbf{u}^2}{c_F^2}, 1 - \frac{3\mathbf{u}^2}{c_F^2}, \frac{u_x}{c_F}, -\frac{u_x}{c_F}, \frac{u_y}{c_F}, -\frac{u_y}{c_F}, \frac{u_x^2 - u_y^2}{c_F^2}, \frac{u_x u_y}{c_F^2}\right)^\mathsf{T}. \tag{29}$$

The body force is incorporated using Guo's forcing scheme [41], written here in moment space as

$$\begin{aligned} \mathbf{G}_F = \Bigg[&0, \frac{6(F_{b,x}u_x + F_{b,y}u_y)}{c_F^2}, -\frac{6(F_{b,x}u_x + F_{b,y}u_y)}{c_F^2}, \frac{F_{b,x}}{c_F}, -\frac{F_{b,x}}{c_F}, \\ &\frac{F_{b,y}}{c_F}, -\frac{F_{b,y}}{c_F}, \frac{2(F_{b,x}u_x - F_{b,y}u_y)}{c_F^2}, \frac{F_{b,y}u_x + F_{b,x}u_y}{c_F^2}\Bigg]^\mathsf{T}. \end{aligned} \tag{30}$$

The relaxation matrix for the flow field is

$$\mathbf{S}^F = \mathrm{diag}\left(0, s_e, s_\epsilon, 0, s_q, 0, s_q, s_\nu, s_\nu\right), \tag{31}$$

where the non-hydrodynamic relaxation rates are fixed at $s_e = 1.2$, $s_\epsilon = 1.4$, and $s_q = 1.2$. The viscous relaxation rate $s_\nu = 1/\tau_F$ is related to the kinematic viscosity by

$$\tau_F = 3\delta t_F \nu + \frac{1}{2}. \tag{32}$$

Furthermore, to suppress the velocity across the diffuse solid-liquid interface, a weighted partial bounce-back treatment is adopted [21, 43], i.e.,

$$f_i^F(\mathbf{x}+\mathbf{e}_i\delta x, t+\delta t_F) = f_{\mathrm{L,mid}} f_i^{F,\dagger}(\mathbf{x},t) + \left(1 - f_{\mathrm{L,mid}}\right) f_{\bar{i}}^{F,\dagger}(\mathbf{x}+\mathbf{e}_i\delta x, t), \tag{33}$$

where $\bar{i}$ denotes the lattice direction opposite to $i$, and $f_{\mathrm{L,mid}} = \left[f_{\mathrm{L}}(\mathbf{x}+\mathbf{e}_i\delta x, t) + f_{\mathrm{L}}(\mathbf{x},t)\right]/2$ is the liquid fraction at the midpoint between the local and neighboring lattice nodes.

### 3.3. Multirate strategy and source-term coupling

The coupled phase, concentration, temperature, and flow equations evolve on distinct physical and numerical time scales associated with interfacial dynamics, solutal diffusion, thermal diffusion, and hydrodynamic transport. More generally, coupled multiphysics systems with strongly separated time scales may involve stiff source terms and temporal inconsistency when strongly coupled quantities are exchanged between asynchronously advanced fields [34, 35]. In the LBM framework, the relaxation time depends on both the transport coefficient and the numerical time interval. Enforcing a single time interval for all four fields may therefore lead to an unfavorable compromise among numerical stability, accuracy, and computational efficiency. Multirate LBM strategies have previously been developed for CDEs [32] and subsequently applied to coupled thermosolutal dendritic-growth simulations [21]. Here, this multirate concept is combined with the unified MRT formulation described above, while a phase-increment-based treatment is introduced for the asynchronous transfer of phase-change source terms.

The minimum time interval among the four fields is taken as the baseline time step, and the field-specific time intervals are defined as

$$\delta t_\chi = N_\chi \delta t_{\mathrm{base}}, \qquad \chi \in \{\phi, U, T, F\}, \tag{34}$$

where $N_\chi$ is a positive integer update factor. The global clock advances with $\delta t_{\mathrm{base}}$, whereas field $\chi$ is advanced only when its scheduled update time is reached. Whenever a coupled field is updated, the latest available macroscopic variables are used.

A direct evaluation of $Q_U \delta t_U$ or $Q_T \delta t_T$ during a coarse scalar update becomes problematic when $\delta t_U$ or $\delta t_T$ is substantially larger than $\delta t_\phi$. Similar source-term coupling issues are well recognized in conventional phase-change computations, where careful treatment of the latent-heat source is required to maintain numerical stability and consistency [44]. In the present multirate setting, several distinct interface displacements may occur between two consecutive scalar updates, and representing their combined phase-change effect through a single coarse-step source evaluation may introduce temporal lag or a concentrated source contribution. To avoid this inconsistency, the transport-related and phase-change contributions are treated separately. The source terms in Eqs. (8) and (12) are decomposed as

$$\begin{aligned} Q_U &= Q_U^{\mathrm{tr}} + Q_U^{\mathrm{pc}}, \quad Q_U^{\mathrm{pc}} = \frac{\left[1+(1-k_c)U\right]\partial_t\phi - 2\nabla\cdot\mathbf{j}_{\mathrm{at}}}{(1+k_c)-(1-k_c)\phi}, \\ Q_T &= Q_T^{\mathrm{tr}} + Q_T^{\mathrm{pc}}, \quad Q_T^{\mathrm{pc}} = \frac{\rho_{\mathrm{L}} L_h}{2(\rho C_p)_{\mathrm{eff}}}\partial_t\phi, \end{aligned} \tag{35}$$

where the superscripts "tr" and "pc" denote the transport-related correction and phase-change contribution, respectively. Accordingly, $Q_U^{\mathrm{tr}}$ and $Q_T^{\mathrm{tr}}$ correspond to the first terms on the right-hand sides of Eqs. (8) and (12), respectively, and are evaluated only during their scheduled concentration or temperature updates using $\delta t_U$ or $\delta t_T$. These terms are retained on the scalar-field time scale because they form part of the corresponding transport operators and should remain discretely consistent with the diffusion and advection updates. In contrast, the phase-change contributions are directly tied to the instantaneous interface evolution through $\partial_t\phi$. They are therefore converted into increments at every phase-field update and immediately transferred to the corresponding scalar fields, so that their temporal integration follows the resolved phase evolution rather than being determined by the coarser scalar update intervals. This treatment can be viewed as a first-order operator splitting between phase-change coupling and scalar transport.

Let $q$ index consecutive phase-field updates. The resolved phase increment over one phase step is

$$\Delta\phi^q = \phi^{q+1} - \phi^q. \tag{36}$$

Rather than explicitly evaluating $\partial_t\phi$ over a coarse scalar interval, the anti-trapping current is integrated over the individual phase-field step, i.e.,

$$\mathbf{J}_{\mathrm{at}}^q = \int_{t_\phi^q}^{t_\phi^{q+1}} \mathbf{j}_{\mathrm{at}}\, dt \approx \frac{W_0}{2\sqrt{2}}\left[1+(1-k_c)U^\star\right]\Delta\phi^q \mathbf{n}^{q+1}, \tag{37}$$

where the superscript $\star$ denotes the latest available scalar state. Approximating the coefficients as constant over an individual phase step yields the corresponding integrated phase-change increments, i.e.,

$$\Delta U_{\rm pc}^{q} = \frac{\left[1 + (1 - k_c)U^{\star}\right]\Delta\phi^{q} - 2\nabla\cdot\mathbf{J}_{\rm at}^{q}}{(1 + k_c) - (1 - k_c)\phi^{q+1}},$$
$$\Delta T_{\rm pc}^{q} = \frac{\rho_{\rm L} L_h}{2(\rho C_p)_{\rm eff}^{q+1}}\Delta\phi^{q}. \tag{38}$$

Because the quantities in Eq. (38) are already integrated over one phase-field interval, no additional factor involving $\delta t_\phi$, $\delta t_U$, or $\delta t_T$ is required. The increments are immediately injected isotropically into the corresponding scalar distributions, i.e.,

$$f_i^U \leftarrow f_i^U + \omega_i \Delta U_{\rm pc}^q, \quad U \leftarrow U + \Delta U_{\rm pc}^q,$$
$$f_i^T \leftarrow f_i^T + \omega_i \Delta T_{\rm pc}^q, \quad T \leftarrow T + \Delta T_{\rm pc}^q. \tag{39}$$

Since $\sum_i \omega_i = 1$ and $\sum_i \omega_i \mathbf{e}_i = 0$, the injection modifies only the zeroth moment by the prescribed scalar increment and introduces no additional first-moment flux. Updating both the distributions and macroscopic variables makes the redistributed solute and released latent heat immediately available to the subsequent phase-field evolution. Likewise, after each scheduled concentration or temperature transport step, the corresponding macroscopic field is reconstructed immediately so that the next phase update uses the latest updated state.

Over any coarse scalar interval containing multiple phase-field updates, the net phase-change contribution is simply the sum of the phase-step increments that have already been transferred, providing a phase-step quadrature of $\int Q_U^{\rm pc}\, dt$ or $\int Q_T^{\rm pc}\, dt$. Consequently, the transferred source amount is determined by the resolved phase evolution rather than directly by the scalar update factors $N_U$ or $N_T$. The phase-field source $Q_\phi$ is evaluated using the latest available $U$ and $T$, without additional temporal averaging. No analogous phase-change increment is introduced into the flow field, whose body force is evaluated from the currently available macroscopic variables.

The scalar gradients are locally reconstructed from the nonequilibrium distributions [21, 45] as

$$\nabla\chi = -\frac{3}{\tau_\chi}\sum_{i\neq 0}\left(f_i^\chi - f_i^{\chi,\rm eq}\right)\mathbf{e}_i, \qquad \chi \in \{\phi, U, T\}. \tag{40}$$

This formulation avoids direct finite-difference evaluation of $\nabla\phi$, $\nabla U$, and $\nabla T$. The divergence of the integrated anti-trapping current, $\nabla\cdot\mathbf{J}_{\rm at}^q$, is evaluated separately using a compact isotropic finite-difference stencil [46].

The overall computational procedure is summarized in Fig. 1. After initialization of the computational parameters, macroscopic fields, and corresponding distribution functions, the global loop advances with $\delta t_{\rm base}$, and each field is updated according to its update factor $N_\chi$. When a phase-field update is scheduled, $Q_\phi$ is evaluated using the latest $U^\star$ and $T^\star$, the phase distribution is advanced, and the resulting $\Delta\phi^q$ is used to construct and immediately inject $\Delta U_{\rm pc}^q$ and $\Delta T_{\rm pc}^q$. When a concentration or temperature update is scheduled, the corresponding transport-related source term is evaluated and the scalar distributions are advanced through MRT collision, streaming, and macroscopic-variable reconstruction. The flow field is updated independently on its own schedule using the latest available coupled variables. This procedure is repeated until the prescribed final time $t_{\rm end}$ is reached.

Thus, the multirate strategy retains phase-change coupling on the phase-field time scale while allowing the concentration, temperature, and flow transport operators to evolve on their respective numerical time scales. The present multirate MRT-LBM framework is implemented on GPUs using *NVIDIA Warp*, and its source code is publicly available in the *DendriteLBM* repository at https://github.com/TheCreatorLiu/DendriteLBM.

## 4. Numerical validation and discussion

Unless otherwise specified, the lattice spacing and baseline time step are set to $\delta x = 1$ and $\delta t_{\rm base} = 1$, respectively, in the following simulations. The four fields are advanced at the same rate by default, i.e., $N_\phi = N_U = N_T = N_F = 1$. Zero-flux boundary conditions, $\partial\phi/\partial n = \partial U/\partial n = \partial T/\partial n = 0$, are imposed on all boundaries for the phase, concentration, and temperature fields. For the flow field, periodic boundary conditions are imposed in the transverse direction, while the inlet-velocity and outflow conditions are applied in the streamwise direction for each forced-convection case. Any case-specific time-step ratios or boundary conditions are stated explicitly below.

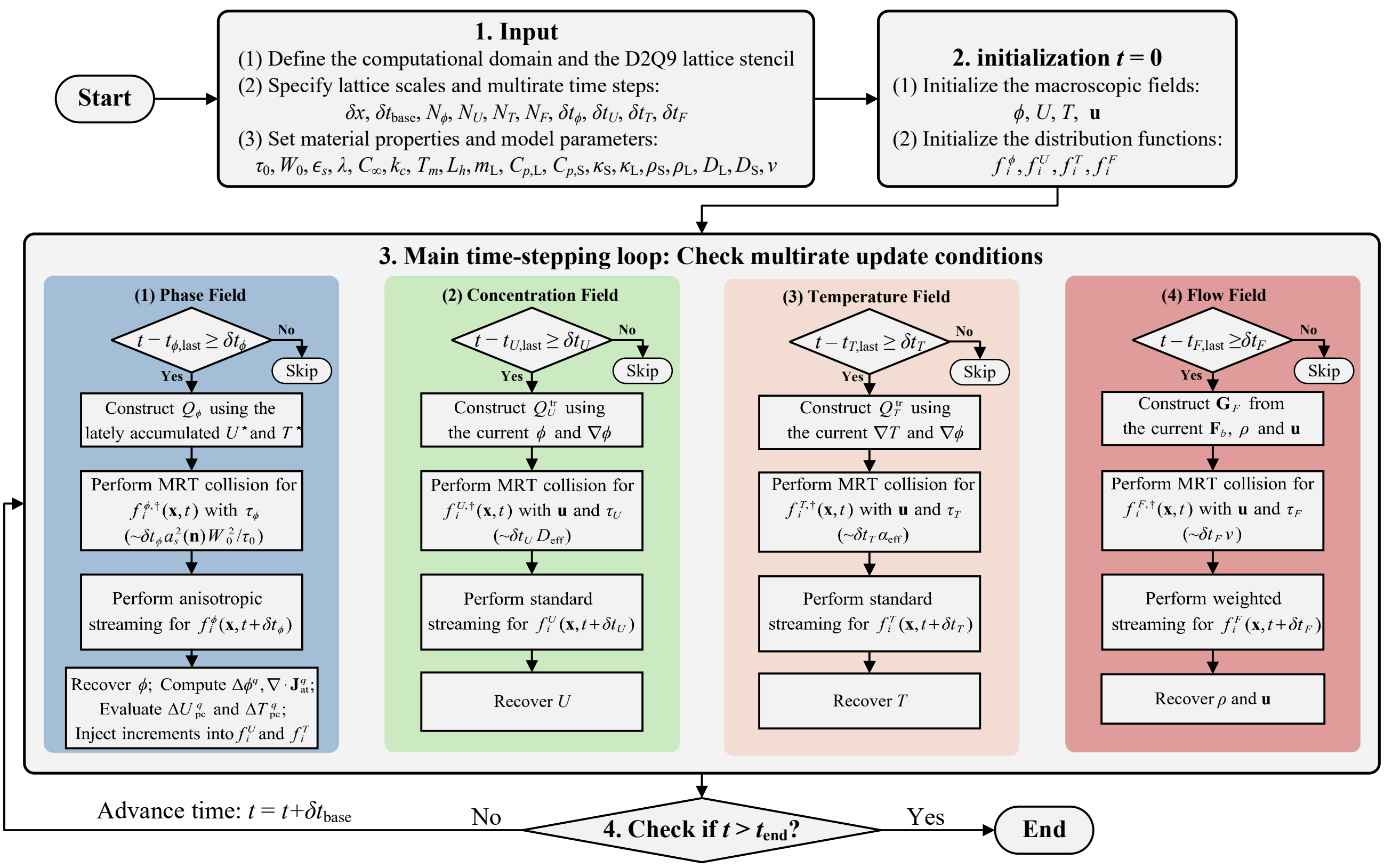

**Figure 1:** Computational workflow of the four-field multirate MRT-LBM, including field-specific updates and phase-increment-based transfer of phase-change contributions.

## 4.1. Thermal dendritic solidification

Thermal dendritic solidification of a pure substance into an undercooled melt is first considered to validate the coupling between the phase and temperature fields. The characteristic length and time scales are set to $W_0 = 2.5\delta x$ and $\tau_0 = 125\delta t_\phi$, respectively. Using the characteristic velocity $W_0/\tau_0$, the thermal Péclet and Prandtl numbers are $Pe = W_0^2/(\alpha_{\text{eff}}\tau_0) = 0.25$ and $Pr = \nu/\alpha_{\text{eff}} = 23.1$. The inlet velocity is $u_{\text{in}} = 0$ for the pure-diffusion case and $u_{\text{in}} = W_0/\tau_0$ for the forced-convection case. A circular seed of radius $R_0 = 10\delta x$ is placed at the center of a square computational domain with $L \times H = 512 \times 512$ lattice nodes and initialized as $\phi(\mathbf{x}, 0) = \tanh[(R_0 - r)/(\sqrt{2}W_0)]$. The effective thermal diffusivity, kinematic viscosity, and latent heat are $\alpha_{\text{eff}} = 0.2$, $\nu = 4.62$, and $L_h = 1.0$, respectively. The anisotropy strength is $\epsilon_s = 0.05$, and the coupling coefficient is $\lambda = a_1 W_0/d_0$, with $d_0 = 0.1385W_0$ and $a_1 = 0.8839$. The initial undercooling is $\theta_0 = -0.55$, and solutal effects are excluded by setting $M_c = 0$. For the forced-convection case, $\delta t_{\text{base}} = 1/15$ is used with $N_F = 1$ and $N_T = 15$. Accordingly, $\delta t_F = 1/15$ gives $\tau_F = 1.424$ according to Eq. (32), placing the flow relaxation time in a numerically favorable range.

Figures 2 and 3 show the phase-field morphology and temperature distribution for thermal dendritic solidification without and with melt flow, respectively, while Fig. 4 compares the corresponding dendrite-tip velocities. In both cases, the $\phi = 0$ interface contours are presented at the same nondimensional times, $t/\tau_0 = 0, 8, 16, 32, 64$, and 128. As shown in Fig. 2(a), the pure-diffusion case preserves fourfold symmetry throughout the evolution, with the four primary branches growing at nearly identical rates along the preferred crystallographic directions. The corresponding isotherms in Fig. 2(b) are likewise symmetric, with closely spaced contours near the dendrite tips reflecting the strong local thermal gradients induced by latent-heat release. Under forced convection, the fourfold symmetry is broken [Fig. 3(a)]: the west branch is enhanced, whereas the east branch is suppressed. Consistently, the isotherms in Fig. 3(b) are compressed on the west side and stretched downstream toward the east, indicating convective redistribution of the thermal boundary layer. Both the dendritic morphologies and temperature distributions under pure diffusion and forced convection are consistent with the corresponding results reported by Zhan et al. [20] and Wang et al. [21].

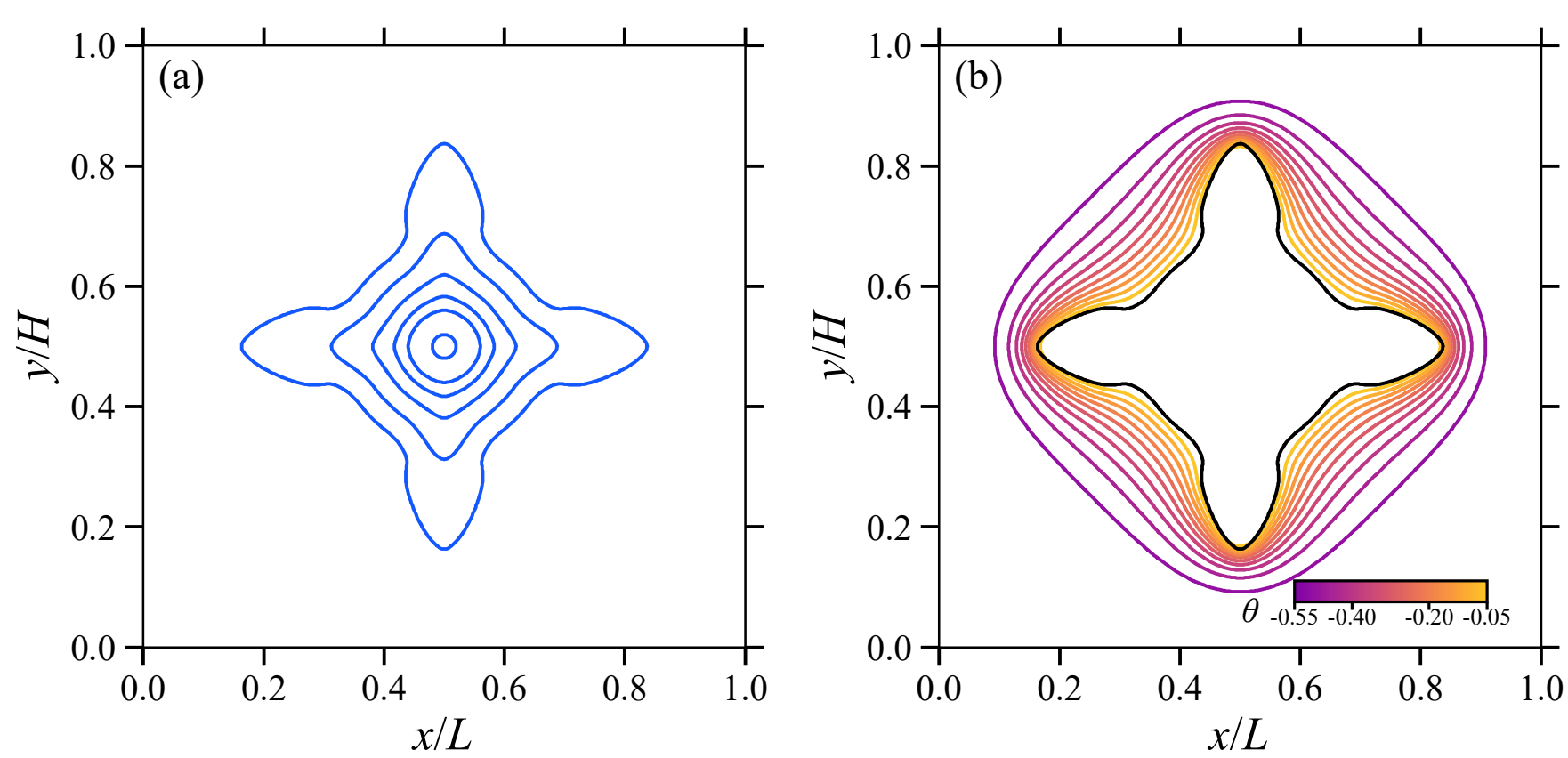

**Figure 2:** Thermal dendritic solidification under pure diffusion with $\theta_0 = -0.55$ and $Pe = 0.25$: (a) $\phi = 0$ interface contours at $t/\tau_0 = 0, 8, 16, 32, 64,$ and 128; and (b) isotherms from $\theta = -0.55$ to $-0.05$ at $t/\tau_0 = 128$.

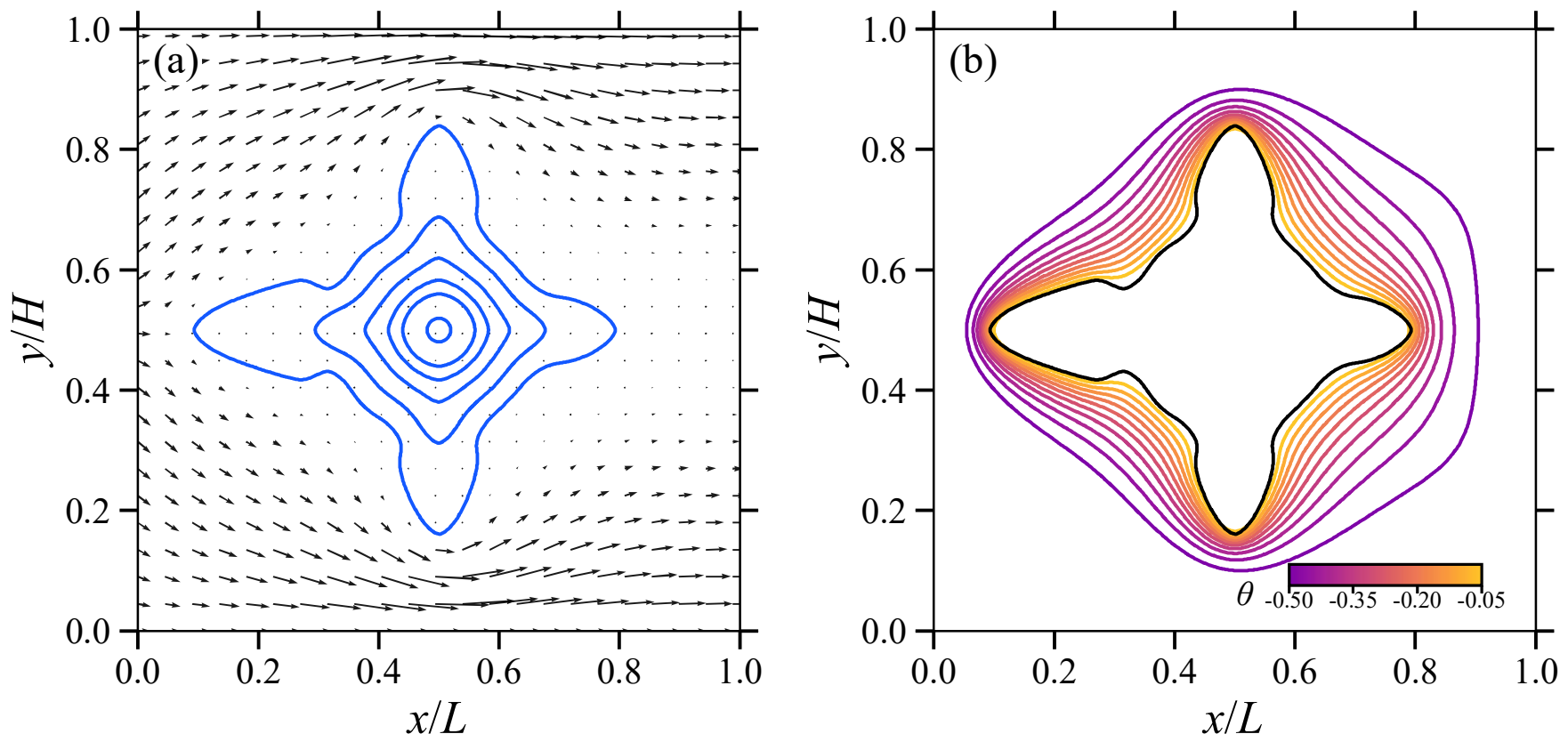

**Figure 3:** Thermal dendritic solidification under forced convection with $\theta_0 = -0.55$, $Pe = 0.25$, and $Pr = 23.1$: (a) $\phi = 0$ interface contours at $t/\tau_0 = 0, 8, 16, 32, 64,$ and 128; and (b) isotherms from $\theta = -0.55$ to $-0.05$ at $t/\tau_0 = 128$.

The tip velocities in Figs. 4(a) and 4(b) are calculated from the axial tip-position histories and nondimensionalized as $v_{\rm tip} d_0/\alpha_{\rm eff}$. In the pure-diffusion case, the four tip velocities are represented by a single averaged curve owing to the fourfold symmetry. Under forced convection, the directional velocities separate after the initial transient, with the west tip growing fastest, the north and south tips exhibiting intermediate velocities, and the east tip growing slowest. Both the temporal evolution and the directional ordering of the predicted tip velocities agree well with the reference results [20, 21]. These comparisons demonstrate that the present multirate MRT-LBM captures the characteristic thermal dendrite-growth dynamics under both diffusive and convective transport. Figure 4(c) further examines the sensitivity to the update factors, $N_T = 15, 30,$ and 45. Despite the threefold variation in $\delta t_T$, the directional tip-velocity histories remain nearly indistinguishable throughout the simulation. This weak sensitivity indicates that the phase-change increments introduced at each phase-field update maintain consistent coupling among the asynchronously advanced fields, supporting the temporal consistency of the multirate treatment over the tested range.

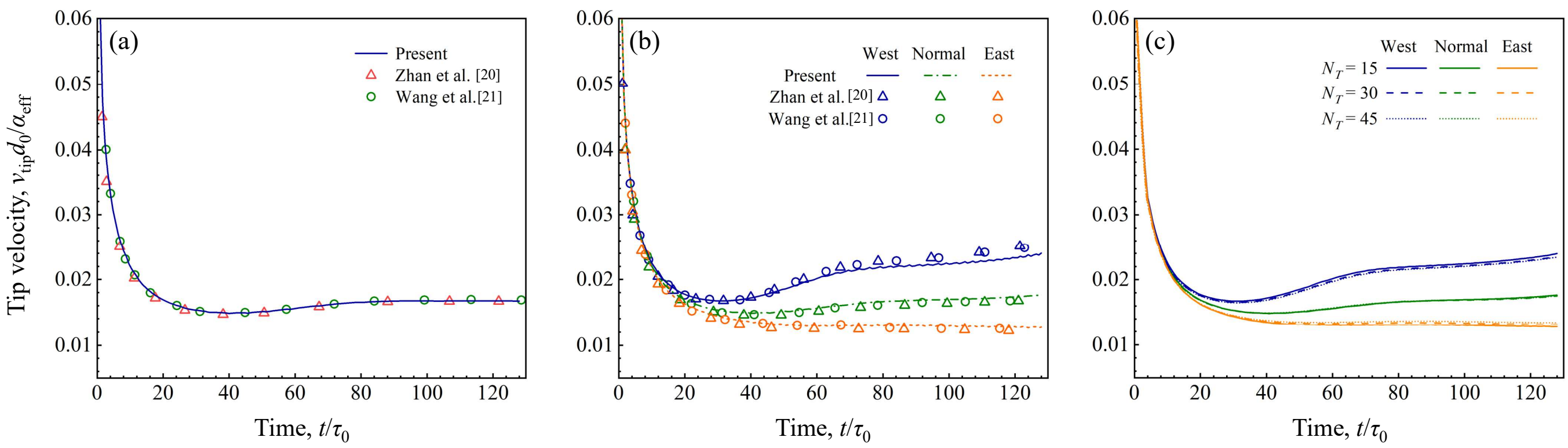

**Figure 4:** Nondimensional tip velocity $v_{\mathrm{tip}}d_0/\alpha_{\mathrm{eff}}$ versus $t/\tau_0$ for thermal dendritic solidification: (a) pure diffusion; (b) directional tip velocities under forced convection; and (c) forced-convection results for temperature-field update factors $N_T = 15, 30,$ and 45.

## 4.2. Solutal dendritic solidification

Isothermal dendritic solidification of a binary system is subsequently considered to validate the coupling between the phase and concentration fields. The initial seed profile is identical to that used in the thermal cases, while the computational domain is enlarged to $L \times H = 1000 \times 1000$ lattice nodes. The phase-field parameters are $W_0 = 2.5\delta x$, $\tau_0 = 50\delta t_\phi$, $\epsilon_s = 0.02$, and $\lambda = a_1 W_0/d_0$, with $d_0 = 0.2762W_0$. Thermal effects are excluded by setting $\theta_0 = 0$, and the initial solutal supersaturation is $U_0 = -0.55$. The solutal coupling parameter and partition coefficient are $M_c = 1$ and $k_c = 0.15$, respectively. The liquid-phase diffusivity is set to $D_{\mathrm{L}} = 0.25$, corresponding to $\widetilde{D}_{\mathrm{L}} = D_{\mathrm{L}}\tau_0/W_0^2 = 2$, while $D_{\mathrm{S}} = 10^{-2}D_{\mathrm{L}}$ is used to suppress solute diffusion in the solid phase. For the forced-convection case, a horizontal inlet velocity of $u_{\mathrm{in}} = W_0/\tau_0$ is prescribed. The baseline time step is set to $\delta t_{\mathrm{base}} = 1/15$, with the corresponding update factors $N_F = 1$ and $N_U = 15$.

Figures 5 and 6 show the phase-field morphology and supersaturation distribution for solutal dendritic solidification without and with melt flow, respectively. In the pure-diffusion case, the dendrite retains fourfold symmetry, while solute rejected at the solid-liquid interface forms a symmetric enrichment layer around the growing branches [Fig. 5]. Under forced convection, the west branch is promoted and the east branch is suppressed, accompanied by compression of the upstream solutal boundary layer and downstream stretching of the supersaturation field [Fig. 6]. These morphological and solutal-field characteristics are consistent with the numerical results reported by Zhan et al. [20] and Wang et al. [21].

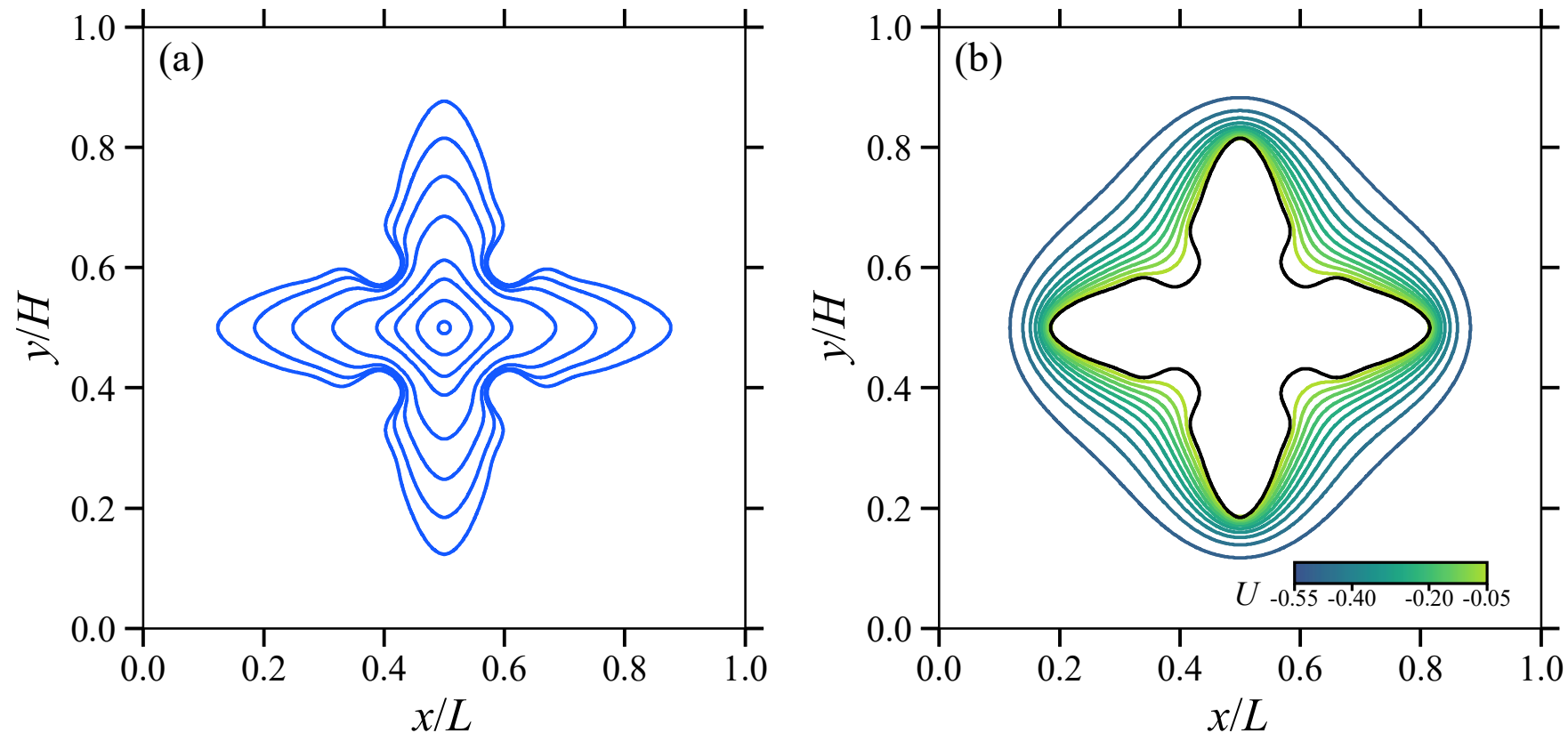

**Figure 5:** Solutal dendritic solidification under pure diffusion with $U_0 = -0.55$ and $\widetilde{D}_{\mathrm{L}} = 2$: (a) $\phi = 0$ interface contours at $t/\tau_0 = 0, 40, 120, 200, 400, 600, 800,$ and 1000; and (b) supersaturation contours from $U = -0.55$ to $-0.05$ at $t/\tau_0 = 1000$.

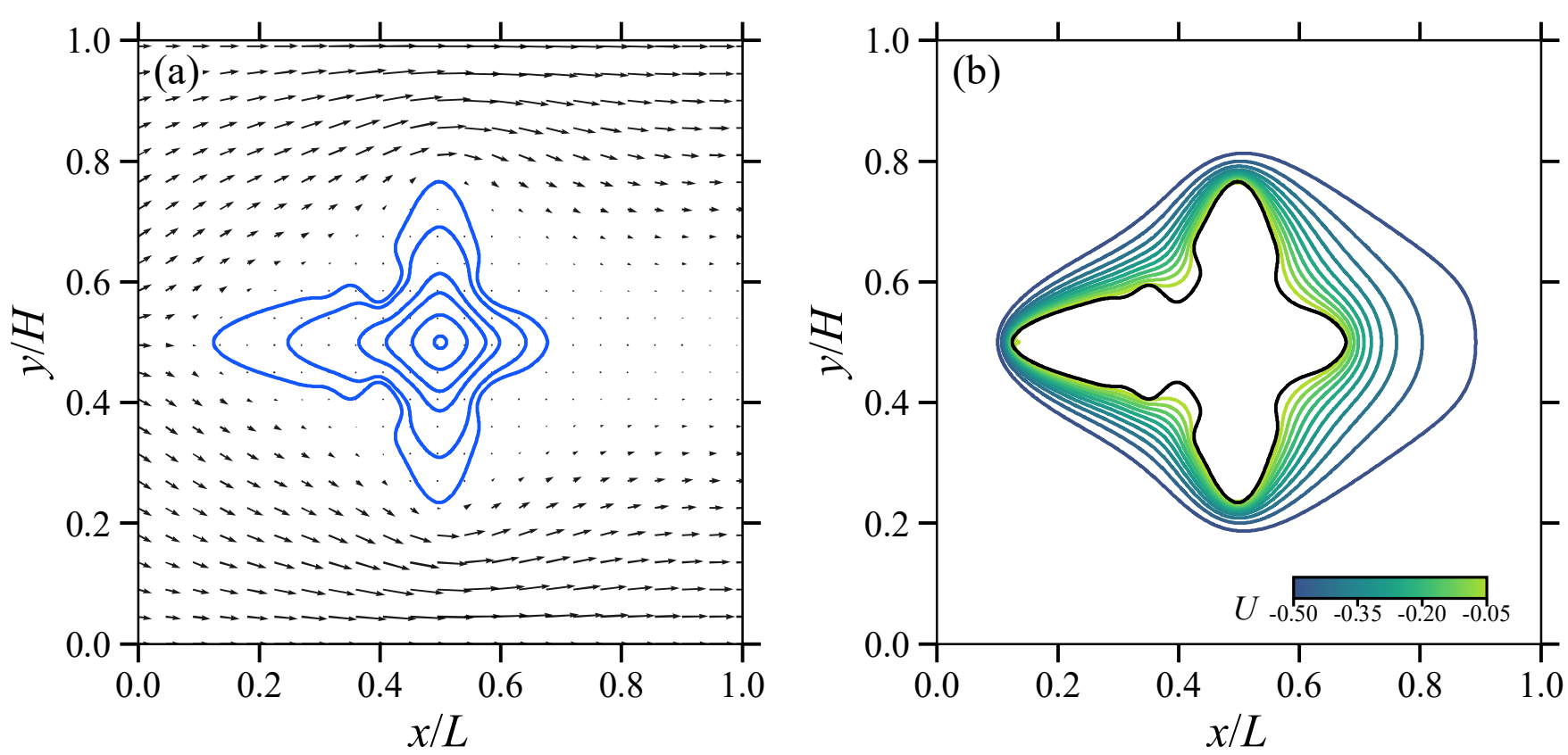

**Figure 6:** Solutal dendritic solidification under forced convection with $U_0 = -0.55$, $\widetilde{D}_\mathrm{L} = 2$, and $u_\mathrm{in} = W_0/\tau_0$: (a) $\phi = 0$ interface contours at $t/\tau_0 = 0, 40, 120, 200, 400,$ and $600$; and (b) supersaturation contours from $U = -0.55$ to $-0.05$ at $t/\tau_0 = 600$.

The tip velocities in Figs. 7(a) and 7(b) are nondimensionalized as $v_\mathrm{tip} d_0/D_\mathrm{L}$. The pure-diffusion result agrees well with the data of Zhan et al. [20] and Wang et al. [21], while the forced-convection case reproduces the directional ordering $v_\mathrm{west} > v_\mathrm{normal} > v_\mathrm{east}$ reported by Zhan et al. [20]. Figure 7(c) further shows that varying the update factor from $N_U = 15$ to 45 produces only minor changes in the directional tip-velocity histories, indicating weak sensitivity of the phase-concentration coupling to the tested update factors.

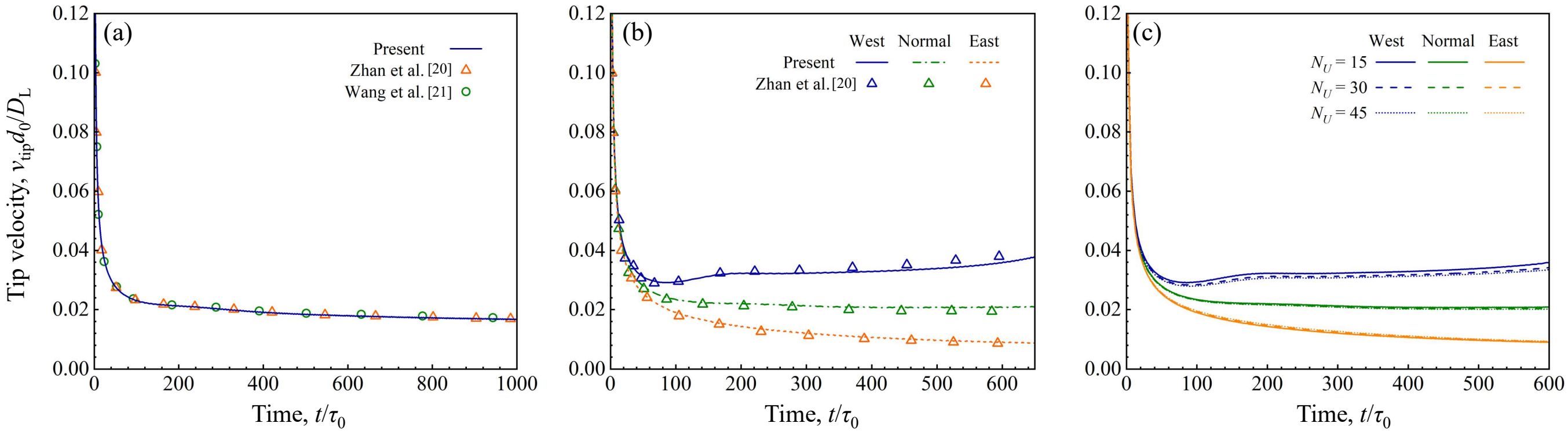

**Figure 7:** Nondimensional tip velocity $v_\mathrm{tip} d_0/D_\mathrm{L}$ versus $t/\tau_0$ for solutal dendritic solidification: (a) pure diffusion; (b) directional tip velocities under forced convection; and (c) forced-convection results for concentration-field update factors $N_U = 15, 30,$ and $45$.

## 4.3. Thermosolutal dendritic solidification

The thermosolutal benchmark originally proposed by Ramirez et al. [11] and later revisited by Wang et al. [21] is considered in this subsection under both pure-diffusion and forced-convection conditions. A computational domain of $L \times H = 1500 \times 1500$ lattice nodes is used, with a circular seed of radius $R_s = 45\delta x$ placed at the domain center. The phase-field parameters are $W_0 = 2\delta x$, $\tau_0 = 1000\delta t_\phi$, $\epsilon_s = 0.02$, $d_0 = 0.554W_0$, and $\lambda = a_1 W_0/d_0 = 1.5955$. The thermodynamic and initial-state parameters are $M_c = 0.1$, $k_c = 0.15$, $U_0 = 0$, and $\theta_0 = -0.55$. A Lewis number of $Le = \alpha_\mathrm{L}/D_\mathrm{L} = 50$ is adopted to examine the multirate treatment under distinctly separated thermal and solutal transport time scales. With $D_\mathrm{L} = 0.004$, the corresponding transport parameters are $\widetilde{D}_\mathrm{L} = D_\mathrm{L}\tau_0/W_0^2 = 1$, $\alpha_\mathrm{L} = LeD_\mathrm{L} = 0.2$, and $D_\mathrm{S} = 10^{-2} D_\mathrm{L} = 4.0 \times 10^{-5}$.

The phase, temperature, and flow fields are updated with $N_\phi = N_T = N_F = 1$, whereas the concentration field is advanced with $N_U = 30$. The resulting concentration relaxation time varies from approximately 0.524 to 0.860 across the solid and liquid phases, while the temperature relaxation time is $\tau_T = 1.1$. The flow field is suppressed in the pure-diffusion case. For forced convection, all other parameters are unchanged, and an inlet velocity of $u_\mathrm{in} = W_0/(5\tau_0)$ is imposed at the left boundary. The corresponding solutal Péclet and Prandtl numbers are $Pe_D = u_\mathrm{in} W_0/D_\mathrm{L} = 0.2$ and $Pr = \nu_\mathrm{L}/\alpha_\mathrm{L} = 23.1$, respectively.

Figure 8 shows the phase, supersaturation, and undercooling fields for the pure-diffusion case at $t^* = tD_\mathrm{L}/d_0^2 = 3500$. The dendrite retains the expected fourfold symmetry, while the coupled transport fields exhibit a clear separation of spatial scales: solute rejection produces a thin enrichment layer adjacent to the solid-liquid interface, whereas the thermal disturbance extends over a much broader region because of the substantially larger thermal diffusivity. This characteristic thermosolutal structure is consistent with the results reported by Ramirez et al. [11] and Wang et al. [21], despite some differences in the detailed branch shapes. The corresponding forced-convection results are shown in Fig. 9. Rather than simply altering the dendrite shape, the imposed flow redistributes the coupled thermal and solutal boundary layers around the growing interface. The upstream enrichment layer becomes compressed, whereas the downstream concentration disturbance is elongated; simultaneously, the flow accelerates around the solid envelope and forms a low-velocity wake downstream. The resulting asymmetric growth and coupled transport characteristics are qualitatively consistent with the forced-convection results reported by Wang et al. [21].

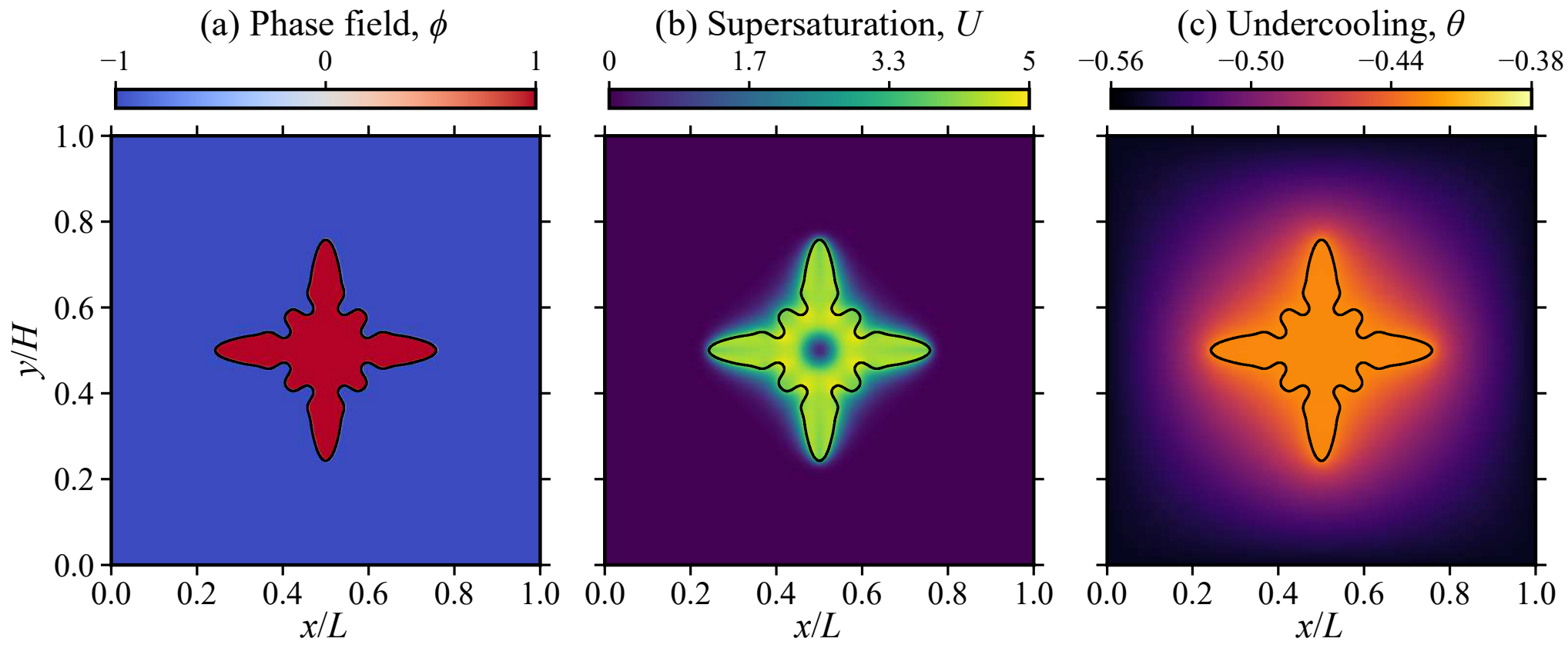

**Figure 8:** Thermosolutal dendritic solidification under pure diffusion at $Le = 50$ and $tD_{\mathrm{L}}/d_0^2 = 3500$: (a) phase field, (b) supersaturation $U$, and (c) undercooling $\theta$. The black curve denotes the $\phi = 0$ interface.

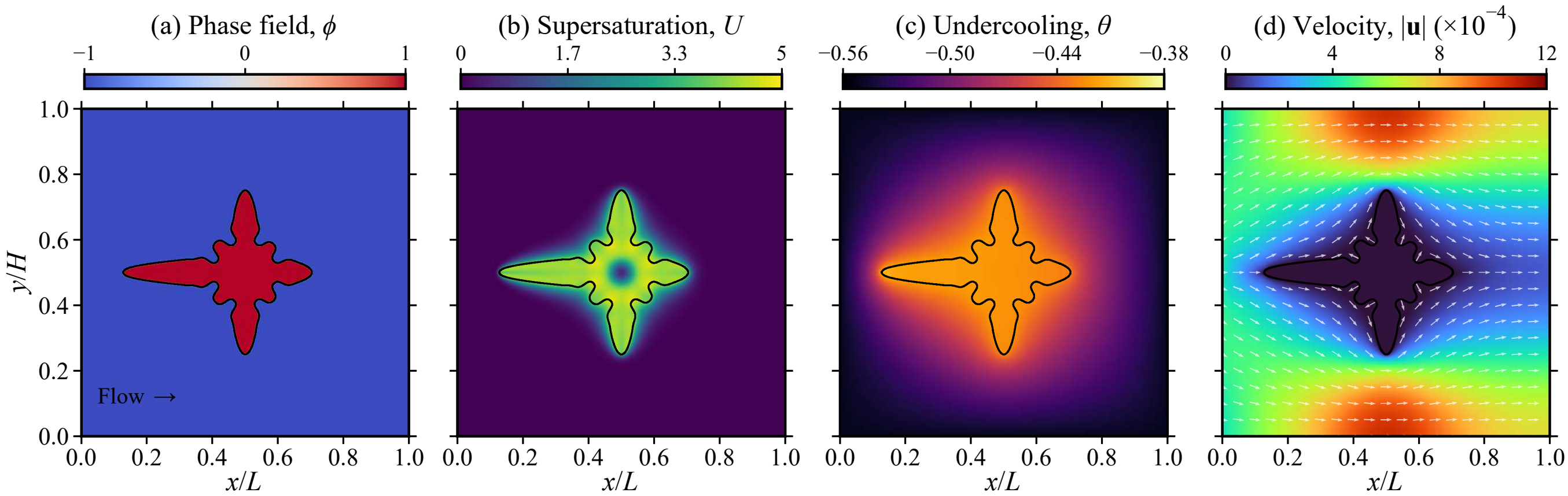

**Figure 9:** Thermosolutal dendritic solidification under forced convection at $Le = 50$, $Pe_D = 0.2$, and $tD_{\mathrm{L}}/d_0^2 = 3500$: (a) phase field, (b) supersaturation $U$, (c) undercooling $\theta$, and (d) velocity magnitude and vectors. The black curve denotes the $\phi = 0$ interface.

In addition to the field distributions, the tip-velocity histories provide a quantitative measure of the growth dynamics. As shown in Fig. 10(a), the present pure-diffusion result and the two reference datasets exhibit the same characteristic evolution, namely a rapid initial decrease in tip velocity followed by a much slower variation. Differences remain in the velocity magnitude and the location of the minimum, while noticeable discrepancies also exist between the two published datasets. These deviations may arise from differences in numerical formulation, spatial resolution, and interface-identification procedures. Nevertheless, the present simulation reproduces the overall temporal behavior of the dendrite-tip velocity. Under forced convection [Fig. 10(b)], the directional tip velocities gradually separate as the dendrite develops, with the upstream west tip growing fastest, the normal tips remaining intermediate, and the downstream east tip growing slowest. The resulting ordering, $v_{\mathrm{west}} > v_{\mathrm{normal}} > v_{\mathrm{east}}$, is consistent with the directional growth trend reported by Wang et al. [21]. Figure 10(b) also examines the sensitivity to the concentration-field update factor by varying $N_U$ from 30 to 50. The directional tip-velocity histories remain very close over the tested range, and the corresponding final interface contours in Fig. 10(c) show only minor differences. These results show that increasing the concentration-field update interval has little influence on either the predicted tip velocity or the final dendritic morphology, confirming weak sensitivity over the tested concentration-field update factors $N_U$. Overall, the multirate treatment provides consistent thermosolutal predictions at $Le = 50$.

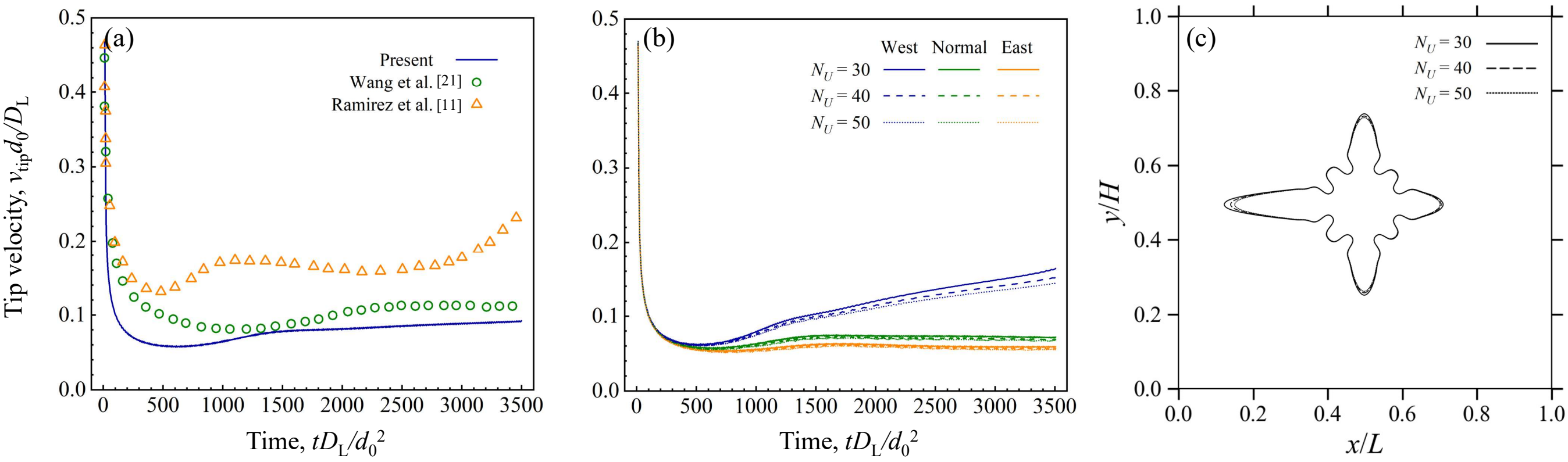

**Figure 10:** Nondimensional tip velocity $v_{tip}d_0/D_L$ versus $tD_L/d_0^2$ for thermosolutal dendritic solidification at $Le = 50$: (a) pure diffusion; (b) directional tip velocities under forced convection at $Pe_D = 0.2$ for concentration-field update factors $N_U = 30, 40,$ and 50; and (c) the corresponding interface contours at $tD_L/d_0^2 = 3500$.

## 4.4. Directional thermosolutal solidification at different Lewis numbers

To further assess the multirate framework using a saline-water configuration and over a wider separation of thermal and solutal transport time scales, directional thermosolutal solidification is simulated over a range of Lewis numbers. The $Le \simeq 100$ case corresponds to directional solidification of saline water, whereas the remaining Lewis numbers are generated by varying the liquid solute diffusivity to provide controlled numerical tests of the multirate treatment. All simulations are performed in a computational domain containing $L \times H = 1000 \times 1000$ lattice nodes, with $W_0 = 2\delta x$ and $\tau_0 = 150\delta t_\phi$. A sinusoidally perturbed solid layer is initialized at the cold bottom boundary as

$$\phi(\mathbf{x}, 0) = \tanh\left[\frac{10\delta x + \delta x \sin(10\pi x/L) - y}{\sqrt{2}W_0}\right], \tag{41}$$

corresponding to a mean interface height of $10\delta x$, a perturbation amplitude of $\delta x$, and a wavelength of $L/5$. The temperature field is initially uniform at $T_b = -20\,°\mathrm{C}$, and the bottom boundary is maintained at this temperature throughout the simulation. The equilibrium melting temperature is $T_m = 0\,°\mathrm{C}$. Melt flow is neglected so that the interface evolution is governed solely by phase change and coupled thermal-solutal diffusion.

The physical-to-lattice conversion factors are $\delta x = 1.0 \times 10^{-7}$ m, $\delta t_{base} = 2.0 \times 10^{-9}$ s, $\delta m = 1.0 \times 10^{-18}$ kg, and $\delta T = 1$ K. The physical properties and corresponding lattice-unit parameters are summarized in Table 1. For the saline-water case, the liquid thermal and solutal diffusivities are $\alpha_L = 1.3 \times 10^{-7}\,\mathrm{m^2\,s^{-1}}$ and $D_L = 1.3 \times 10^{-9}\,\mathrm{m^2\,s^{-1}}$, respectively, giving $Le \simeq 100$. To systematically increase the separation between the thermal and solutal transport time scales, only $D_L$ is varied, while all other parameters are kept unchanged, i.e.,

$$D_L = \left\{1.3 \times 10^{-7},\ 1.3 \times 10^{-8},\ 1.3 \times 10^{-9},\ 1.3 \times 10^{-10}\right\}\ \mathrm{m^2\,s^{-1}}, \qquad Le = \frac{\alpha_L}{D_L} \simeq \{1,\ 10,\ 100,\ 1000\}. \tag{42}$$

Accordingly, the concentration-field update factors are set to $N_U = 1$, 10, 100, and 1000 for $Le = 1$, 10, 100, and 1000, respectively, while $N_\phi = N_T = 1$ is maintained in all cases. Despite the three-order-of-magnitude variation in $D_L$, this choice keeps the liquid-phase concentration relaxation time nearly constant at $\tau_{U,L} = 0.578$. The resulting configuration therefore isolates the effect of increasing thermal-solutal timescale separation without simultaneously driving the concentration relaxation time toward an unfavorable numerical range.

Figure 11 compares the concentration and temperature fields at the common physical time $t = 10^7\delta t_\phi = 20$ ms, while the complete temporal evolution is provided in Supplementary Video S1. At $Le = 1$, $D_L$ is comparable to $\alpha_L$, and the rejected solute is rapidly redistributed before a pronounced solutal boundary layer can develop. Together with capillary smoothing, this rapid diffusion suppresses the initial sinusoidal perturbation and leaves the interface nearly planar. As $Le$ increases, the decreasing solute diffusivity progressively confines solute enrichment to the advancing interface and promotes morphological instability. At $Le = 10$, several broad columnar cells emerge, separated by solute-rich channels and isolated brine pockets. For the saline-water case at $Le = 100$, the thinner solutal boundary layer is accompanied by reduced primary spacing and the formation of a dense array of slender, vertically oriented dendrites with pronounced interdendritic solute enrichment. This morphology is qualitatively consistent with the ice-crystal structures observed experimentally by Yuan et al. [6], particularly the vertically aligned dendrites and solute-rich interdendritic channels. At $Le = 1000$, solute becomes even more strongly confined to the interface, while tip splitting and branch competition become more pronounced. Although the overall growth front remains relatively uniform, the internal solid-liquid interface becomes increasingly complex.

**Table 1.** Physical and lattice parameters for saline-water directional solidification.

| Quantity and symbol | Physical value | Lattice-unit value |
|---|---|---|
| Latent heat, $L_h$ | $3.34 \times 10^5$ J kg$^{-1}$ | 133.6 |
| Specific heat, $C_{p,\mathrm{L}}$ | 4191 J kg$^{-1}$ K$^{-1}$ | 1.676 |
| Specific heat, $C_{p,\mathrm{S}}$ | 2090 J kg$^{-1}$ K$^{-1}$ | 0.836 |
| Thermal conductivity, $\kappa_\mathrm{L}$ | 0.5457 W m$^{-1}$ K$^{-1}$ | 0.0437 |
| Thermal conductivity, $\kappa_\mathrm{S}$ | 2.22 W m$^{-1}$ K$^{-1}$ | 0.178 |
| Thermal diffusivity, $\alpha_\mathrm{L}$ | $1.3 \times 10^{-7}$ m$^2$ s$^{-1}$ | 0.026 |
| Thermal diffusivity, $\alpha_\mathrm{S}$ | $1.06 \times 10^{-6}$ m$^2$ s$^{-1}$ | 0.212 |
| Partition coefficient, $k_c$ | 0.001 | 0.001 |
| Far-field salt fraction, $C_\infty$ | 0.03 | 0.03 |
| Liquidus slope, $m_\mathrm{L}$ | −57 K per unit mass fraction | −57 |
| Solute diffusivity, $D_\mathrm{L}$ ($Le \simeq 100$) | $1.3 \times 10^{-9}$ m$^2$ s$^{-1}$ | $2.6 \times 10^{-4}$ |
| Gibbs-Thomson coefficient, $\Gamma$ | $2.68 \times 10^{-8}$ K m | 0.268 |
| Anisotropy strength, $\epsilon_s$ | 0.03 | 0.03 |
| Capillary length, $d_0 = \dfrac{\Gamma}{\lvert m_\mathrm{L}\rvert(1-k_c)C_\infty}$ | $1.57 \times 10^{-8}$ m | 0.157 |
| Coupling coefficient, $\lambda = a_1 W_0/d_0$ | 11.27 | 11.27 |
| Solutal coupling, $M_c = \dfrac{-m_\mathrm{L}(1-k_c)C_\infty}{L_h/C_{p,\mathrm{L}}}$ | 0.0214 | 0.0214 |

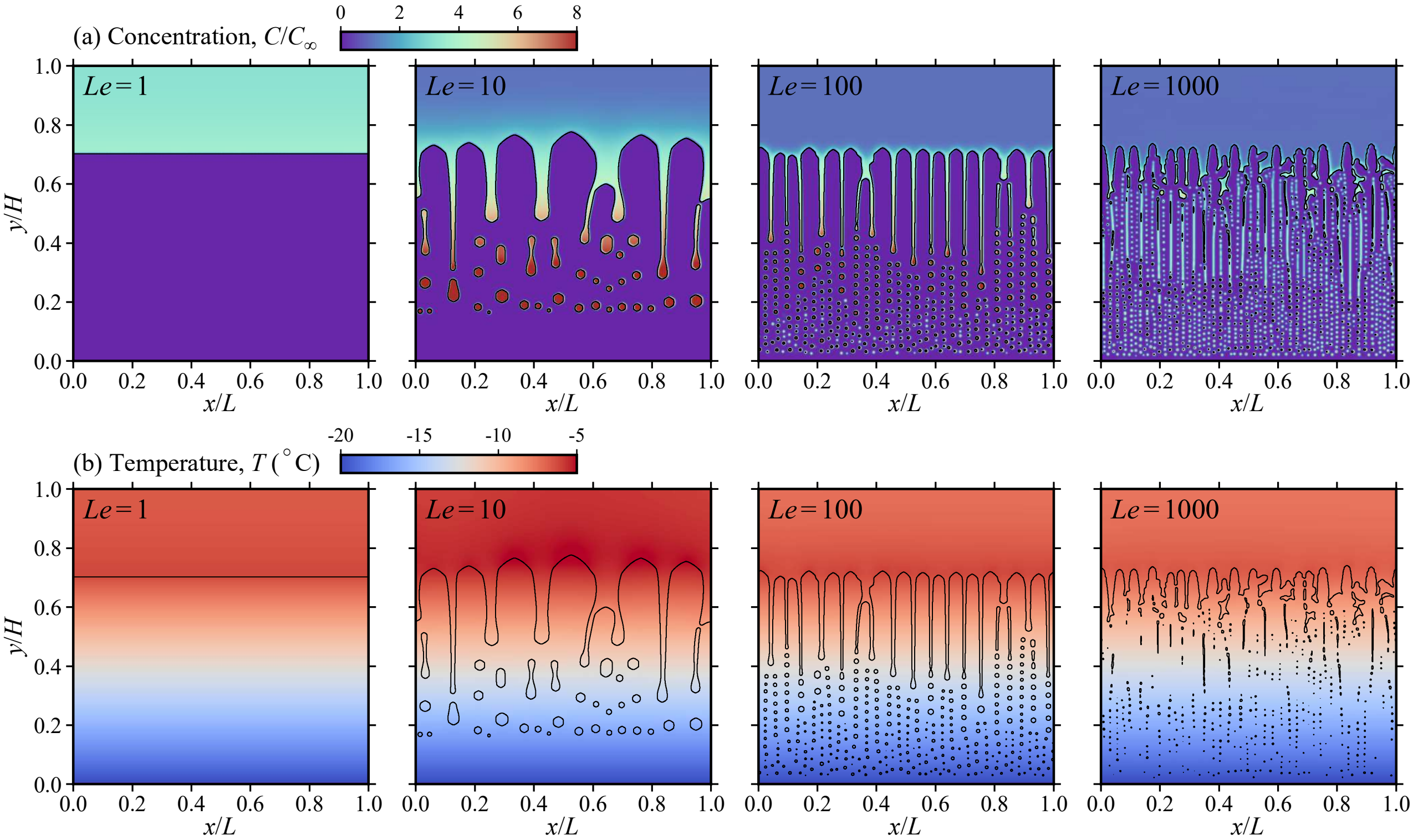

**Figure 11:** Directional thermosolutal solidification at $t = 20$ ms for $Le = 1$, 10, 100, and 1000: (a) normalized concentration $C/C_\infty$ and (b) temperature $T$. The $Le \simeq 100$ case corresponds to the saline-water parameters listed in Table 1. The black curve denotes the $\phi = 0$ interface.

To directly assess the phase-change coupling treatment introduced in Section 3.3, a source-coupling ablation is performed using the directional-solidification configuration of Fig. 11. As described by Eq. (35), the phase-change contribution is separated from the transport-related correction through a first-order operator splitting. In the proposed *instantaneous coupling*, each phase-change increment is transferred immediately after the corresponding phase-field update according to Eq. (39), while the concentration transport operator remains on its scheduled coarse time scale. For comparison, *accumulated coupling* stores the phase-change increments generated between two consecutive concentration updates and injects their sum only at the subsequent coarse update. All other model parameters and numerical settings are kept identical. The comparison is performed for $Le = 1$, 10, and 100, corresponding to $N_U = 1$, 10, and 100, respectively. The accumulated-coupling case at $Le = 1000$ cannot be advanced stably and is therefore excluded from the quantitative comparison.

To quantify the coupling response, two local diagnostics and one global conservation measure are monitored. The maximum phase-change-induced increment is defined as $S_\phi(t) = \max_{\mathbf{x}\in\Omega} |\Delta U^{\mathrm{pc}}(\mathbf{x}, t)|$, while the maximum concentration-field update jump is defined as $J_U(t) = \max_{\mathbf{x}\in\Omega} |U^{\mathrm{post}}(\mathbf{x}, t) - U^{\mathrm{pre}}(\mathbf{x}, t)|$, where $U^{\mathrm{pre}}$ and $U^{\mathrm{post}}$ denote the concentration field immediately before and after the phase-change contribution is applied, respectively. Thus, $S_\phi$ measures the magnitude of the transferred phase-change contribution, whereas $J_U$ quantifies the resulting instantaneous response of the concentration field. The total solute inventory is evaluated as $M_C(t) = \int_\Omega C(\mathbf{x}, t)\,\mathrm{d}\Omega$, with the relative deviation $R_M(t) = [M_C(t) - M_C(0)]/M_C(0)$. Together, $S_\phi$, $J_U$, and $R_M$ characterize the local source pulse, its immediate effect on the concentration field, and the global solute balance, respectively. Figure 12 shows that the difference between the two coupling treatments becomes increasingly pronounced as the concentration-field update factor $N_U$ increases. The upper panels present the temporal evolution of $S_\phi$ and $J_U$, while the lower panels show the corresponding $R_M$.

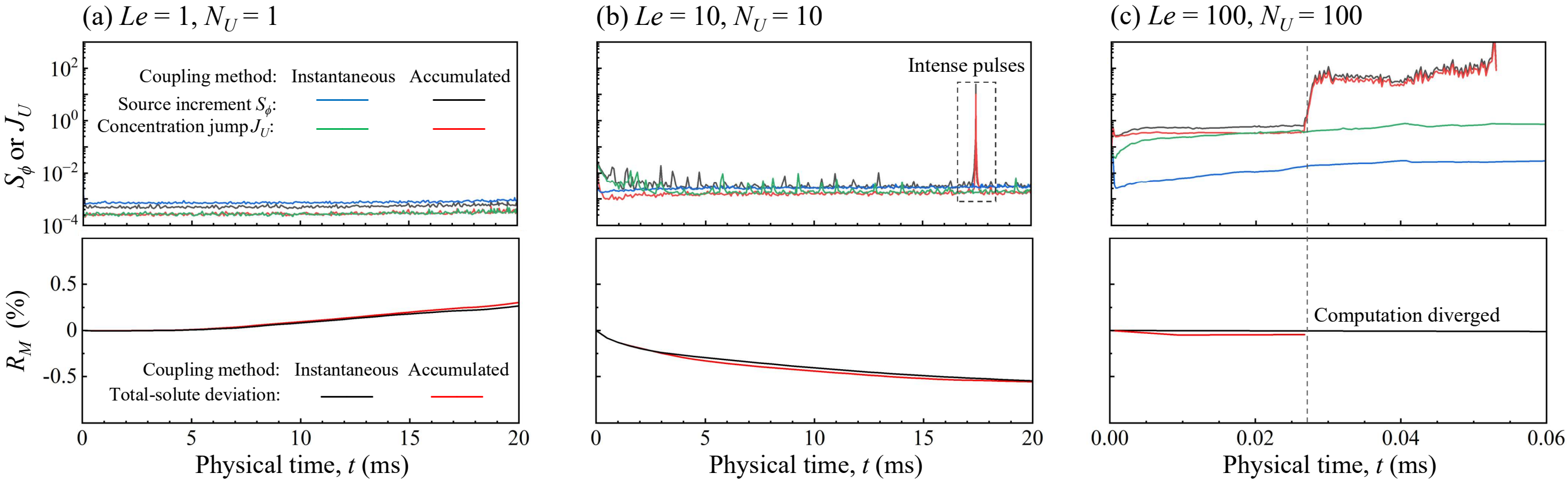

**Figure 12:** Source-coupling ablation for directional thermosolutal solidification: (a) $Le = 1$, (b) $Le = 10$, and (c) $Le = 100$. The upper panels show the temporal evolution of the maximum phase-change-induced increment $S_\phi$ and concentration-field update jump $J_U$ for instantaneous and accumulated coupling, while the lower panels show the corresponding relative deviation of the total solute inventory $R_M$.

At $Le = 1$, where $N_U = 1$, every phase-field update coincides with a concentration-field update, and the two treatments are therefore nearly identical. Both $S_\phi$ and $J_U$ remain at relatively low levels, approximately $10^{-4}$–$10^{-3}$, without pronounced pulse-like oscillations. The final $R_M$ values are +0.27% and +0.30% for instantaneous and accumulated coupling, respectively. At $Le = 10$, the accumulated treatment produces substantially larger peaks in both $S_\phi$ and $J_U$, with increasingly evident pulse-like oscillations at later times. Nevertheless, both calculations remain stable to $t = 20$ ms, and the final $R_M$ values remain close at −0.55% and −0.56%. Thus, the local source pulses become apparent before any significant difference emerges in the global solute balance. The difference becomes critical at $Le = 100$. Under accumulated coupling, $S_\phi$ and $J_U$ develop increasingly strong intermittent pulses, followed by loss of numerical stability and non-finite values at $t = 0.0532$ ms. In contrast, instantaneous coupling remains stable to $t = 20$ ms, and the instantaneous treatment also remains stable for the more strongly separated case at $Le = 1000$. These results show that delaying the phase-change transfer progressively concentrates the interfacial contribution into coarse-step source pulses as the temporal separation increases.

Overall, the ablation directly supports the source-term treatment proposed in Section 3.3. The first-order splitting allows scalar transport to retain its coarse multirate schedule, while the immediate transfer in Eq. (39) prevents phase-change contributions from accumulating into large source pulses. This treatment therefore substantially improves numerical stability under strongly separated phase and solutal transport time scales.

## 5. Conclusions

A unified multirate MRT-LBM has been developed for thermal, solutal, and thermosolutal dendritic solidification with or without melt flow. The coupled fields share a common D2Q9 moment-space framework and advance with independent update factors. To maintain temporal consistency under asynchronous evolution, the transport-related and phase-change contributions in the concentration and temperature equations are treated separately. At each phase-field update, the resolved phase increment $\Delta\phi$ is used to construct and immediately transfer the corresponding solutal and latent-heat increments, allowing the phase-change contributions to follow the resolved interface evolution rather than being accumulated into coarser scalar-field updates. The temperature formulation further incorporates phase-dependent thermal conductivity and volumetric heat capacity to account for solid-liquid property contrasts across the diffuse interface.

The proposed framework was validated using thermal, solutal, and thermosolutal dendritic benchmarks under pure diffusion and forced convection. The predicted dendritic morphologies and tip-velocity trends are consistent with previous numerical results and reproduce the characteristic directional growth induced by forced convection. Over the tested range, the thermal and solutal results show only weak sensitivity to the scalar-field update factors. For the thermosolutal case at Lewis number $Le = 50$, the predicted tip velocities and final dendritic morphologies remain close as the concentration-field update factor $N_U$ is increased from 30 to 50. Further directional solidification simulations over $Le = 1$–$1000$ capture the progressive localization of solute near the interface and the transition from an almost planar interface to cellular and strongly branched structures. For the saline-water case at $Le \simeq 100$, the model produces vertically aligned dendrites and solute-rich interdendritic channels in qualitative agreement with experimental observations. Despite the three-order-of-magnitude variation in solute diffusivity and concentration-field update factor, the overall solute balance remains well maintained. Source-coupling ablation further shows that, as the time-scale separation increases, accumulating phase-change contributions over coarse scalar-field steps produces increasingly strong local source pulses and is followed by loss of numerical stability at larger $Le$. In contrast, phase-step instantaneous transfer remains stable over the tested conditions. These results support the temporal consistency, numerical stability, and applicability of the proposed multirate coupling strategy under strongly separated transport time scales.

The present study employs first-order operator splitting and has so far been validated only in two-dimensional configurations. Future work may explore higher-order temporal coupling, adaptive selection of field-specific update factors, and extension of the framework to three-dimensional dendritic simulations.

## Acknowledgments

This work was supported by the National Natural Science Foundation of China (Grant No. 52376061). The authors thank Dr. Shaotong Fu of NVIDIA Beijing for his technical support in GPU-accelerated computing.

## Code availability

The DendriteLBM source code for the benchmarks in Sections 4.1–4.3 is available under the MIT License at https://github.com/TheCreatorLiu/DendriteLBM.